\documentclass{IJCAS}

\usepackage{url}
\usepackage{array,tabularx}
\usepackage{multicol}
\usepackage{amsmath,amssymb}
\usepackage{booktabs}
\usepackage{multirow}
\usepackage{makecell}
\usepackage{siunitx}
\usepackage{graphicx}
\usepackage{tikz}
\usetikzlibrary{arrows.meta,positioning,calc,shapes.geometric,fit}

\providecommand{\citep}[1]{\cite{#1}}
\providecommand{\sep}{,\ }

\newcolumntype{L}{l}

\journalvolumn{}
\journalnumber{}
\journalyear{}
\setarticlestartpagenumber{1}

\allowdisplaybreaks

\makeatletter
\def\ps@plain{%
  \def\@oddhead{}%
  \def\@evenhead{}%
  \def\@oddfoot{\hfil\thepage\hfil}%
  \let\@evenfoot\@oddfoot}
\def\ps@headings{%
  \def\@oddhead{}%
  \def\@evenhead{}%
  \def\@oddfoot{\hfil\thepage\hfil}%
  \let\@evenfoot\@oddfoot}
\def\ps@myheadings{%
  \def\@oddhead{}%
  \def\@evenhead{}%
  \def\@oddfoot{\hfil\thepage\hfil}%
  \let\@evenfoot\@oddfoot}
\def\ps@IJCAS{%
  \def\@oddhead{}%
  \def\@evenhead{}%
  \def\@oddfoot{\hfil\thepage\hfil}%
  \let\@evenfoot\@oddfoot}
\def\ps@ijcas{%
  \def\@oddhead{}%
  \def\@evenhead{}%
  \def\@oddfoot{\hfil\thepage\hfil}%
  \let\@evenfoot\@oddfoot}
\AtBeginDocument{\pagestyle{plain}\thispagestyle{plain}}
\makeatother

\begin{document}

\title{Deep Koopman risk-preview supervised LTV-MPC for direct yaw moment control of distributed drive electric vehicles}

\author{Wenjie Wang, Hao Chen, Ran Shu, Kyoungseok Han, and Hongyu Shu*}

\begin{abstract}
Always-on direct yaw moment control (DYC) improves vehicle stability during critical maneuvers but can introduce unnecessary interventions under low-risk conditions. This paper proposes a Koopman risk-gated linear time-varying model predictive control (KRG-LTV-MPC) framework for low-intervention yaw stability assistance. Instead of replacing the physics-based execution model with a fully data-driven control predictor, this framework separates Koopman-based phase-risk preview from safety-critical execution. A Deep Koopman model predicts the nominal evolution of the sideslip--yaw rate phase risk to determine whether the constrained quadratic programming (QP) problem should be solved or skipped at each sampling time. When the gate is active, the LTV-MPC layer calculates the additional yaw moment; otherwise, the QP is skipped and the previously commanded moment is tapered to zero under a bounded-rate rule. Event-level shadow-mode evaluation shows that the Koopman predictor provides positive warning lead times of 0.19--0.28~s under low-friction and friction-transition conditions, whereas the LTV predictor gives delayed warnings. Under closed-loop low-friction conditions, KRG-LTV-MPC reduces the cumulative yaw-moment intervention by 43.9\% relative to LTV-MPC and solves the QP for only 41.0\% of the samples while maintaining vehicle stability within the phase plane. These results support the use of Koopman phase-risk information as an intelligent supervisory layer for low-intervention DYC.
\end{abstract}


\maketitle

\makeAuthorInformation{
Wenjie Wang is with the College of Mechanical and Vehicle Engineering, Chongqing University, Chongqing, 400044, China, and the Department of Automotive Engineering, Hanyang University, Seoul, 04763, South Korea (e-mail: wwj@stu.cqu.edu.cn). Hao Chen and Hongyu Shu are with the College of Mechanical and Vehicle Engineering, Chongqing University, Chongqing, 400044, China (e-mail: chen.h@cqu.edu.cn; shycqu@cqu.edu.cn). Ran Shu is with the School of Vehicle Engineering, Chongqing University of Technology, Chongqing, 401135, China (e-mail: ranshu@cqut.edu.cn). Kyoungseok Han is with the Department of Automotive Engineering, Hanyang University, Seoul, 04763, South Korea (e-mail: kyoungsh@hanyang.ac.kr). 

* Corresponding author.}

\thispagestyle{plain}
\pagestyle{plain}

\section{Introduction}
\label{sec:introduction}
Distributed drive electric vehicles (DDEVs) provide an effective platform for direct yaw moment control (DYC), because wheel torques can be regulated rapidly and independently~\citep{zhang2025review}. By generating asymmetric longitudinal tire forces between the left and right wheels, DYC produces an additional yaw moment that suppresses excessive sideslip and yaw rate deviation during emergency steering, low-friction and friction-varying maneuvers~\citep{HUA2025106437,jin2026vehicle,WANG2025109600}. Model predictive control (MPC) is well suited to yaw stability control because yaw moment limits, rate constraints, stability envelope constraints, and motor torque limits can all be handled explicitly over a finite prediction horizon~\citep{wu2025coordinated,Zhang01022023}. However, conventional linear time-varying MPC (LTV-MPC)-based DYC is often implemented as an always-on stabilizing controller~\citep{guo2018coordinated}. Although this design is effective near the handling limit, it may introduce unnecessary yaw moment intervention during moderate-risk or recovery phases, increasing actuator usage and interfering with driver steering input. Therefore, for DYC of DDEVs, the control problem is not only how to compute the corrective yaw moment but also when the yaw moment optimization should be activated.

The stability envelope~\citep{efremov2024vehicle} and the phase plane methods~\citep{zhu2023survey} provide a geometric basis to judge whether the vehicle is approaching the handling limit. In particular, the sideslip--yaw rate phase plane can characterize the evolution of vehicle stability and distinguish recoverable states from divergent responses near the tire-road friction limit~\citep{wang2025coordinated,tristano2026analytical}. Nevertheless, a risk decision based only on the measured current state is essentially reactive. It may activate DYC too late under fast friction transitions or nonlinear tire saturation, whereas a conservative threshold may trigger unnecessary intervention in low-risk phases. This motivates a preview-oriented supervisory decision that anticipates the future evolution of the phase plane trajectory before the vehicle state reaches the stability boundary.

Linear two-degree-of-freedom (2-DOF) and LTV predictors are computationally efficient, physically interpretable, and directly expose the actuation channel from the additional yaw moment to the sideslip--yaw rate response, making them suitable for the constrained execution layer of a DYC controller~\citep{cheng2020model}. However, such linear predictions can become optimistic or delayed when the vehicle operates in nonlinear tire saturation regions~\citep{wang2025hybrid}, under rapidly varying road friction~\citep{xu2026stochastic}, or during combined longitudinal--lateral maneuvers~\citep{zhang2023coordinated}. Data-driven vehicle models have also been integrated with MPC-based torque vectoring to compensate for unmodeled dynamics and changing operating conditions~\citep{kim2025data}. In the present work, this motivates retaining the LTV model for safety-critical yaw moment execution while introducing a nonlinear preview model for the supervisory risk decision.

The Koopman operator theory~\citep{koopman1931hamiltonian} provides a data-driven route to nonlinear prediction in a form compatible with finite-horizon control. By lifting the nonlinear dynamics into a higher-dimensional observable space, the nonlinear evolution can be approximated by a linear predictor. Finite-dimensional approximations have been constructed using dynamic mode decomposition~\citep{rowley2017model}, extended dynamic mode decomposition~\citep{ghosh2024koopman}, and neural-network-based Deep Koopman models~\citep{yeung2019learning}. Koopman linear predictors have been integrated with MPC~\citep{korda2018linear} and have been applied to  vehicle modeling~\citep{kim2023koopman}, trajectory tracking~\citep{wu2026adaptive,zhang2026physics,zuo2025model}, path following~\citep{ren2026data,wang2026data}, and robotic motion control~\citep{mamakoukas2021derivative,otto2021koopman,wang2022improved}. More closely related to vehicle yaw control, a Koopman predictor has been used directly as a plant model inside a predictive torque vector optimizer~\citep{vsvec2023predictive,vsvec2025optimizing}, and robust tube-based formulations have been proposed to handle uncertainty of the Koopman model inside the optimizer~\citep{zhang2022robust}. Meanwhile, a deep Koopman model has also been integrated with control barrier functions as an external safety command governor~\citep{chen2024deep}. However, in these studies, the Koopman model either replaces the physical execution model inside the optimizer or acts as an external cascaded filtering layer that continuously modifies the control input. In contrast, using a Deep Koopman predictor as a supervisory phase-risk preview module to selectively activate a physically constrained DYC execution layer for DDEVs has not been explored sufficiently. 

The present work therefore adopts a model-role-separated architecture. The yaw moment execution layer requires a physically interpretable and constraint-consistent mapping from yaw moment to the sideslip--yaw rate response, whereas the control sensitivity of a learned predictor cannot be guaranteed to remain reliable across unseen operating conditions. Building on this motivation, this paper proposes a Deep Koopman risk-preview supervised LTV-MPC framework for DYC of DDEVs. From an intelligent system perspective, the framework converts Koopman-predicted vehicle responses into phase-risk information for solver activation, while retaining the LTV-MPC layer as the physically constrained yaw moment execution model. Specifically, the Deep Koopman model predicts the future nominal phase-risk evolution, and a composite phase-risk index converts the predicted trajectory into a gate signal. The LTV-MPC QP is solved only when the gate is active; otherwise, the QP is skipped and the previous yaw moment is tapered to zero at a bounded rate. The resulting solver-gated controller is termed Koopman risk-gated LTV-MPC (KRG-LTV-MPC). 

The main contributions of this paper are summarized as follows:
\begin{itemize}
    \item A model-role-separated Deep Koopman risk-preview supervised LTV-MPC architecture is developed for DYC of DDEVs, decoupling Koopman-based phase-risk preview from physically constrained yaw moment execution.
    \item A composite phase-risk index is developed to evaluate sideslip magnitude, predicted divergence away from the stable region, and whether the yaw rate direction is favorable for reducing the sideslip angle.
    \item A solver-gated KRG-LTV-MPC controller is formulated to reduce unnecessary DYC intervention by activating the LTV-MPC QP only when the Koopman-based risk gate is active.
    \item Event-level shadow-mode evaluation, gate-source ablation, and computation time analysis are conducted to assess early phase-risk detection, the stability--intervention trade-off, and solver usage reduction.
\end{itemize}

The remainder of this paper is organized as follows. Section~\ref{sec:problem} establishes theoretical preliminaries. Section~\ref{sec:risk_gated} presents the proposed KRG-LTV-MPC controller. Section~\ref{sec:experiments} presents the validation experiments and results. Section~\ref{sec:conclusion} concludes this paper.

\section{Preliminaries}
\label{sec:problem}

\subsection{2-DOF vehicle model}
\label{subsec:states_envelope}

\begin{figure}
	\centering
	\includegraphics[width=\columnwidth]{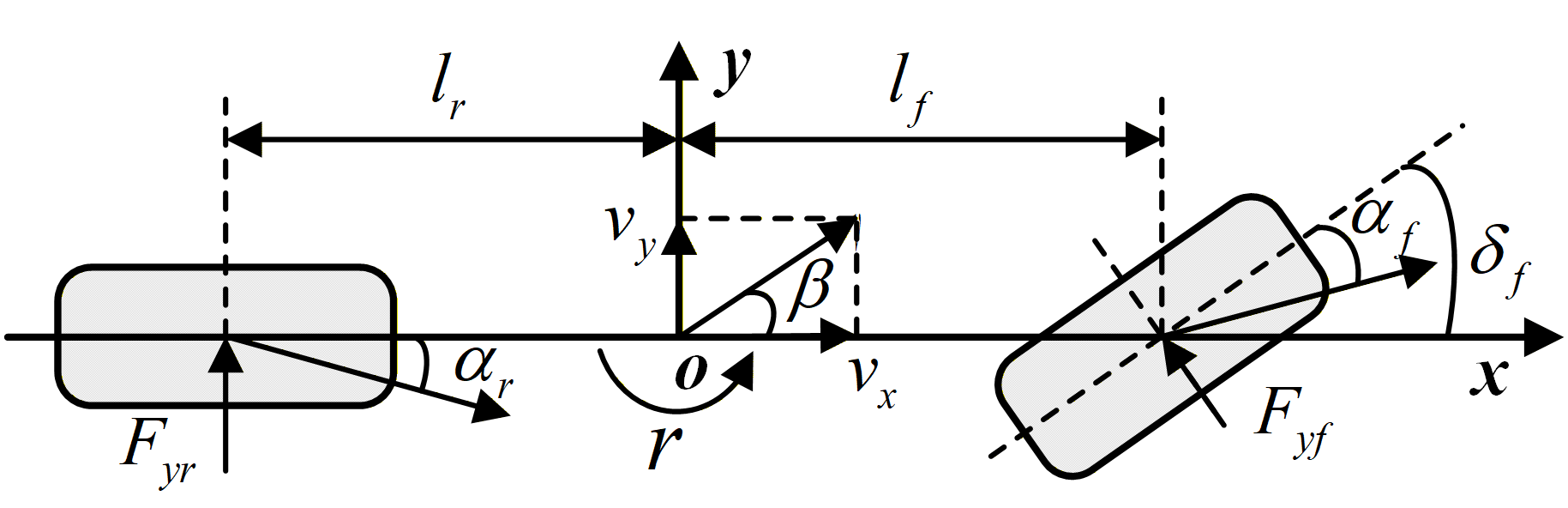}
	\caption{Planar 2-DOF vehicle model.}
	\label{fig:vehicle_schematic}
\end{figure}

The execution-layer controller is based on a planar 2-DOF model, as shown in Fig.~\ref{fig:vehicle_schematic}. The longitudinal velocity $v_x$ is treated as a measured scheduling variable, and the additional yaw moment $M_z$ is treated as the upper-level DYC input. The force and moment balance equations are
\begin{equation}
	\left\{
	\begin{aligned}
		m (\dot v_y + v_x r) &= F_{yf} + F_{yr}, \\
		I_z \dot r &= l_f F_{yf} - l_r F_{yr} + M_z,
	\end{aligned}
	\right.
	\label{eq:2DOF_vehicle_dynamics}
\end{equation}
where $m$ is the vehicle mass, $v_y$ is the lateral velocity, $r$ is the yaw rate, $I_z$ is the yaw moment of inertia, $l_f$ and $l_r$ are the distances from the center of gravity to the front and rear axles, and $F_{yf}$ and $F_{yr}$ are the front and rear lateral tire forces.

The sideslip angle is defined from the measured velocity components as $\beta=\arctan(v_y/v_x)$. For model derivation, the small-angle approximation $v_y\simeq v_x\beta$ is used. The front and rear tire slip angles are then written as
\begin{equation}
	\left\{
	\begin{aligned}
		\alpha_f &= \beta+\frac{l_f r}{v_x}-\delta_f, \\
		\alpha_r &= \beta-\frac{l_r r}{v_x},
	\end{aligned}
	\right.
	\label{eq:tire_slip_angles_final}
\end{equation}
where $\delta_f$ is the front wheel steering angle.

With positive cornering stiffness $C_{\alpha f}$ and $C_{\alpha r}$, the lateral tire forces are approximated by the linear model
\begin{equation}
	\left\{
	\begin{aligned}
		F_{yf} &=-C_{\alpha f}\alpha_f
		=-C_{\alpha f}\left(\beta+\frac{l_f r}{v_x}-\delta_f\right), \\
		F_{yr} &=-C_{\alpha r}\alpha_r
		=-C_{\alpha r}\left(\beta-\frac{l_r r}{v_x}\right).
	\end{aligned}
	\right.
	\label{eq:linear_tire_forces_final}
\end{equation}
Substituting Eq.~\eqref{eq:linear_tire_forces_final} into Eq.~\eqref{eq:2DOF_vehicle_dynamics} gives the continuous-time LTV model
\begin{equation}
	\dot{\mathbf{x}}^{\mathrm{LTV}}=\mathbf{A}_c \mathbf{x}^{\mathrm{LTV}} + \mathbf{B}_{\delta}\delta_f + \mathbf{B}_{M}M_z,
	\label{eq:ltv_cont_model_final}
\end{equation}
where
\begin{equation}
\begin{split}
	\mathbf{A}_c &= \begin{bmatrix} -\dfrac{C_{\alpha f}+C_{\alpha r}}{m v_x} & \dfrac{l_r C_{\alpha r}-l_f C_{\alpha f}}{m v_x^2}-1 \\[2mm] \dfrac{l_r C_{\alpha r}-l_f C_{\alpha f}}{I_z} & -\dfrac{l_f^2 C_{\alpha f}+l_r^2 C_{\alpha r}}{I_z v_x} \end{bmatrix}, \\[2mm]
	\mathbf{x}^{\mathrm{LTV}} &= \begin{bmatrix} \beta\\[1mm] r \end{bmatrix}, \quad
	\mathbf{B}_{\delta}=\begin{bmatrix} \dfrac{C_{\alpha f}}{m v_x}\\[2mm] \dfrac{l_fC_{\alpha f}}{I_z} \end{bmatrix},\quad
	\mathbf{B}_{M}=\begin{bmatrix} 0\\[2mm] \dfrac{1}{I_z} \end{bmatrix}.
\end{split}
\label{eq:matrices_ltv_final}
\end{equation}

To support the discrete-time MPC formulation, Eq.~\eqref{eq:ltv_cont_model_final} is discretized by the forward Euler method with a sampling time $T_s$ as
\begin{equation} 
	\mathbf{x}_{k+1}^{\mathrm{LTV}} = \mathbf{A}_d \mathbf{x}_k^{\mathrm{LTV}} + \mathbf{B}_{d,\delta}\delta_{f,k} + \mathbf{B}_{d,M}M_{z,k}, 
	\label{eq:ltv_discrete_final} 
\end{equation}
where
\begin{equation}
	\mathbf{A}_d=\mathbf{I}_2+T_s\mathbf{A}_c,
	\qquad
	\mathbf{B}_{d,\delta}=T_s\mathbf{B}_{\delta},
	\qquad
	\mathbf{B}_{d,M}=T_s\mathbf{B}_{M}.
	\label{eq:ltv_discrete_matrices_final}
\end{equation}
The stability envelope is defined as
\begin{equation}
	|\beta|\le \beta_{\lim},\qquad |r|\le r_{\lim},
	\label{eq:basic_envelope_final}
\end{equation}
with
\begin{equation}
	\beta_{\lim}=\arctan(0.02 \mu g),
	\qquad
    r_{\lim}=0.85\frac{\mu g}{v_x},
	\label{eq:lim_final}
\end{equation}
where $\mu$ is the road friction coefficient and $g$ is the gravitational acceleration. The same envelope is used for phase-risk normalization and for the LTV-MPC state constraints.

\subsection{Deep Koopman model for nominal risk preview}
\label{subsec:koopman_predictor_final}
\begin{figure}
	\centering
	\includegraphics[width=\columnwidth]{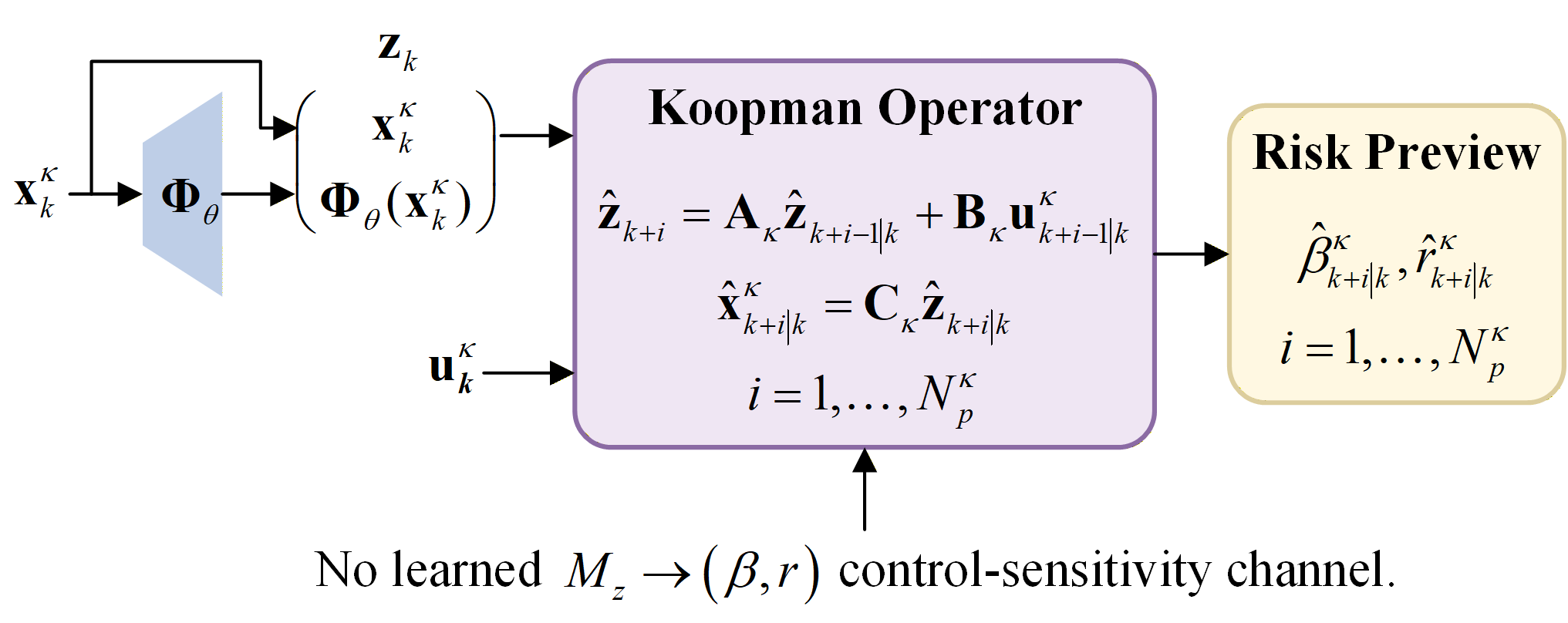}
	\caption{Deep Koopman model for nominal risk preview.}
	\label{fig:koopman_architecture}
\end{figure}

The Deep Koopman predictor used in this study is based on a previously developed vehicle dynamics modeling framework~\citep{wang2026adaptivekoopman}. In the present paper, it serves as a nominal Koopman-based risk-preview model rather than as a prediction model inside the LTV-MPC QP. As shown in Fig.~\ref{fig:koopman_architecture}, the Koopman model predicts the future sideslip--yaw rate response under the total torque demand and steering input, and the predicted sequence is subsequently converted into the phase-risk index used by the solver gate.

The structure of the observable network $\boldsymbol{\Phi}_{\theta}$ was chosen as $[3~128~128~128~16]$. The model was trained offline using CarSim data from randomized figure-of-eight maneuvers on a high-friction road ($\mu=0.85$). The training loss combines reconstruction, lifted linearity, multi-step prediction, and physics-informed tire-force equilibrium residuals. In that prior modeling study~\citep{wang2026adaptivekoopman}, this model achieved open-loop prediction RMSEs of $0.0868~\mathrm{m/s}$ for lateral velocity and $0.0363~\mathrm{rad/s}$ for yaw rate over a 5-s horizon under extreme steering maneuvers.

The predictor uses the physical state
\begin{equation}
	\mathbf{x}_k^{\mathcal{K}}
	=
	\begin{bmatrix}
		v_{x,k} & v_{y,k} & r_k
	\end{bmatrix}^{\mathrm T}
	\label{eq:koopman_state_final}
\end{equation}
and the lifted state
\begin{equation}
	\mathbf{z}_k
	=
	\mathbf{g}_{\theta}\left(\mathbf{x}_k^{\mathcal{K}}\right)
	=
	\begin{bmatrix}
		\left(\mathbf{x}_k^{\mathcal{K}}\right)^{\mathrm T} &
		\boldsymbol{\Phi}_{\theta}\left(\mathbf{x}_k^{\mathcal{K}}\right)^{\mathrm T}
	\end{bmatrix}^{\mathrm T}
	\in\mathbb{R}^{19}.
	\label{eq:koopman_lifted_state_final}
\end{equation}
The lifted linear dynamics are
\begin{equation}	
	\mathbf{z}_{k+1}
	=
	\mathbf{A}_{\mathcal{K}}\mathbf{z}_k
	+
	\mathbf{B}_{\mathcal{K}}\mathbf{u}_k^{\mathcal{K}},
	\qquad
	\mathbf{u}_{k}^{\mathcal{K}}
	=
	\begin{bmatrix}
		T_{\mathrm{tot},k} & \delta_{\mathrm{sw},k}
	\end{bmatrix}^{\mathrm T},
	\label{eq:koopman_nominal_final}
\end{equation}
where $\mathbf{A}_{\mathcal{K}}\in\mathbb{R}^{19\times19}$ and
$\mathbf{B}_{\mathcal{K}}\in\mathbb{R}^{19\times2}$ are the lifted system and input matrices, respectively. $T_{\mathrm{tot}}$ is the total torque demand and $\delta_{\mathrm{sw}}$ is the steering wheel angle. 

Because the physical state is explicitly embedded in the lifted vector, the predicted vehicle state can be recovered by the linear projection
\begin{equation}
	\hat{\mathbf{x}}_{k}^{\mathcal{K}}
	=
	\mathbf{C}_{\mathcal{K}}\hat{\mathbf{z}}_k,
	\qquad
	\mathbf{C}_{\mathcal{K}}
	=
	\begin{bmatrix}
		\mathbf{I}_3 & \mathbf{0}_{3\times16}
	\end{bmatrix}.
	\label{eq:koopman_projection_final}
\end{equation}

At each sampling time, the Koopman model is rolled out over the preview horizon $N_p^{\mathcal{K}}$ with the initial condition
$\hat{\mathbf{z}}_{k|k}=\mathbf{z}_k$ as
\begin{equation}
	\left\{
	\begin{aligned}
		\hat{\mathbf{z}}_{k+i|k}
		&=
		\mathbf{A}_{\mathcal{K}}\hat{\mathbf{z}}_{k+i-1|k}
		+
		\mathbf{B}_{\mathcal{K}}\mathbf{u}_{k+i-1|k}^{\mathcal{K}},\\
		\hat{\mathbf{x}}_{k+i|k}^{\mathcal{K}}
		&=
		\mathbf{C}_{\mathcal{K}}\hat{\mathbf{z}}_{k+i|k},
	\end{aligned}
	\right.
	\qquad
	i=1,\ldots,N_p^{\mathcal{K}}.
	\label{eq:koopman_preview_rollout_final}
\end{equation}
The predicted sideslip angle is computed as
\begin{equation}
	\hat{\beta}_{k+i|k}^{\mathcal{K}}
	=
	\arctan
	\left(
	\frac{\hat v_{y,k+i|k}^{\mathcal{K}}}
	{\hat v_{x,k+i|k}^{\mathcal{K}}}
	\right),
	\quad
	i=1,\ldots,N_p^{\mathcal{K}}.
	\label{eq:koopman_beta_r_preview_final}
\end{equation}

The input vector $\mathbf{u}_{k}^{\mathcal{K}}$ does not include an independently trained yaw moment channel. Therefore, the Koopman model is used only to generate the nominal preview sequence $\{ \hat\beta_{k+i|k}^{\mathcal{K}}, \hat r_{k+i|k}^{\mathcal{K}} \}_{i=1}^{N_p^{\mathcal{K}}}$, which is converted into the phase-risk index in Section~\ref{subsec:risk_preview}.
\subsection{Baseline LTV-MPC}
\label{subsec:ltv_mpc_final}
The LTV-MPC baseline is constructed from the discretized 2-DOF model in Eq.~\eqref{eq:ltv_discrete_final}. Over the prediction horizon, the stacked output relation is
\begin{equation}
    Y^{\mathrm{LTV}}=Y_0^{\mathrm{LTV}}+\Theta^{\mathrm{LTV}}U,
    \label{eq:ltv_stacked_final}
\end{equation}
where the free-response vector $Y_0^{\mathrm{LTV}}\in\mathbb{R}^{2N_p}$ and the forced-response matrix $\Theta^{\mathrm{LTV}}\in\mathbb{R}^{2N_p\times N_c}$ are constructed from the LTV prediction matrices as
\begin{equation}
\begin{split}
    Y_0^{\mathrm{LTV}} &= \mathcal{A}_d\,\mathbf{x}_k^{\mathrm{LTV}}+\mathcal{B}_{d,\delta}\boldsymbol{\delta}_k, \\
    \Theta^{\mathrm{LTV}} &=
    \begin{bmatrix}
        \mathbf{B}_{d,M} & \mathbf{0} & \cdots & \mathbf{0}\\
        \mathbf{A}_d\mathbf{B}_{d,M} & \mathbf{B}_{d,M} & \cdots & \mathbf{0}\\
        \vdots & \vdots & \ddots & \vdots\\
        \mathbf{A}_d^{N_p-1}\mathbf{B}_{d,M} & \cdots & \cdots & \mathbf{A}_d^{N_p-N_c}\mathbf{B}_{d,M}
    \end{bmatrix},
\end{split}
\label{eq:ltv_stacked_matrices}
\end{equation}
with $\mathcal{A}_d=[\mathbf{A}_d;\,\mathbf{A}_d^2;\,\ldots;\,\mathbf{A}_d^{N_p}]\in\mathbb{R}^{2N_p\times 2}$ and $\mathcal{B}_{d,\delta}$ the corresponding steering accumulation matrix. Here $\boldsymbol{\delta}_k$ denotes the future steering sequence over the prediction horizon, which is treated as a measured disturbance at each sampling time using the current steering measurement held constant. The vector $Y^{\mathrm{LTV}}=[\beta_{k+1|k},r_{k+1|k},\ldots,\beta_{k+N_p|k},r_{k+N_p|k}]^\top$ stacks the predicted sideslip angle and yaw rate, and the control input sequence is
\begin{equation}
    U=[M_{z,k|k},M_{z,k+1|k},\ldots,M_{z,k+N_c-1|k}]^\top.
\end{equation}

The yaw rate reference is computed from the driver's steering input and saturated by the friction-adaptive limit, while the sideslip angle reference is set to zero for stability regulation.
\begin{equation}
	\left\{
	\begin{aligned}
		r_{\mathrm{ref}} &= \min \left\{ \left| \frac{v_x}{L(1 + K_s v_x^2)} \delta_f \right|, \left| 0.85 \frac{\mu g}{v_x} \right| \right\} \operatorname{sgn}(\delta_f), \\
		\beta_{\mathrm{ref}} &= 0,
	\end{aligned}
	\right.
	\label{eq:reference_states_final}
\end{equation}
where $L=l_f+l_r$ is the wheelbase and $K_s$ is the vehicle stability factor, given by
\begin{equation}
	K_s = \frac{m}{L^2} \left( \frac{l_r}{C_{\alpha f}} -\frac{l_f}{C_{\alpha r}} \right).
	\label{eq:Ks_final}
\end{equation}

The control increment is defined as $\Delta M_{z,k+i|k} = M_{z,k+i|k} - M_{z,k+i-1|k}$, with $M_{z,k-1|k}=M_z(k-1)$ denoting the yaw moment applied at the previous sampling time. The LTV-MPC solves the following QP:
\begin{equation}
\begin{aligned}
J_{\mathrm{LTV}}=&\sum_{i=1}^{N_p}
\left[Q_\beta(\beta_{k+i|k}-\beta_{\mathrm{ref}})^2
+Q_r(r_{k+i|k}-r_{\mathrm{ref},k+i})^2\right]\\
&+\sum_{i=0}^{N_c-1}\left(R_M M_{z,k+i|k}^2+R_{\Delta M}\Delta M_{z,k+i|k}^2\right)
+\rho_{\epsilon}\epsilon^2,
\end{aligned}
\label{eq:ltv_mpc_cost_final}
\end{equation}
subject to
\begin{equation}
	\left\{
	\begin{aligned}
		|\beta_{k+i|k}| &\le \beta_{\lim} + \epsilon, & i &= 1,\ldots,N_p,\\
		|r_{k+i|k}|     &\le r_{\lim}     + \epsilon, & i &= 1,\ldots,N_p,\\
		|M_{z,k+i|k}|   &\le M_{z,\max},              & i &= 0,\ldots,N_c-1,\\
		|\Delta M_{z,k+i|k}| &\le \Delta M_{z,\max},  & i &= 0,\ldots,N_c-1,\\
		0 &\le \epsilon \le \epsilon_{\max}.
	\end{aligned}
	\right.
	\label{eq:mpc_constraints}
\end{equation}
The prediction and control horizons are set to $N_p=N_c=20$. KRG-LTV-MPC retains the same LTV-MPC prediction model, QP structure, and constraints, while the Koopman risk gate determines only when the QP is solved.
\subsection{Torque allocation strategy}
\label{subsec:torque_allocation}
The target DDEV is equipped with four independently driven in-wheel motors. Since torque allocation optimization is not the focus of this paper, a direct left-right differential allocation is used to realize the upper-level total torque demand $T_{\mathrm{tot}}$ and additional yaw moment command $M_z$. The nominal base torque is
\begin{equation}
	T_{\mathrm{base}}=\frac{T_{\mathrm{tot}}}{4}.
\end{equation}
The front and rear differential torque components are
\begin{equation}
	\Delta T_f=\frac{1}{2}\frac{M_z R_w}{B_f\cos\delta_f},
	\qquad
	\Delta T_r=\frac{1}{2}\frac{M_z R_w}{B_r},
	\label{eq:differential_torque}
\end{equation}
where $B_f$ and $B_r$ are the front and rear track widths, and $R_w$ is the effective wheel radius. The unsaturated wheel torque commands are
\begin{equation}
	\left\{
	\begin{aligned}
		T_{\mathrm{fl}} &= T_{\mathrm{base}}-\Delta T_f,\qquad &T_{\mathrm{fr}} = T_{\mathrm{base}}+\Delta T_f, \\
		T_{\mathrm{rl}} &= T_{\mathrm{base}}-\Delta T_r,\qquad &T_{\mathrm{rr}} = T_{\mathrm{base}}+\Delta T_r.
	\end{aligned}
	\right.
	\label{eq:torque_allocation}
\end{equation}

The actuator limits are configured according to a ProteanDrive Pd18-class in-wheel motor. For each wheel $j\in\{\mathrm{fl},\mathrm{fr},\mathrm{rl},\mathrm{rr}\}$, the available torque limit is
\begin{equation}
	T_{\lim,j}=\begin{cases}
		\min\left(T_{\max},\dfrac{P_{\max}}{\max(|\omega_j|,\omega_{\min})}\right), & |\omega_j|\le \omega_{\max},\\[2mm]
		0, & |\omega_j|>\omega_{\max},
	\end{cases}
	\label{eq:motor_limit}
\end{equation}
where $T_{\max}$, $P_{\max}$, and $\omega_{\max}$ are the peak torque, peak power, and maximum motor speed, respectively. $\omega_{\min}$ is a small positive threshold to prevent division by zero. The final wheel torque command satisfies
\begin{equation}
	-T_{\lim,j}\le T_j\le T_{\lim,j}.
\end{equation}

\section{Koopman risk-gated LTV-MPC}
\label{sec:risk_gated}
\begin{figure*}
	\centering
	\includegraphics{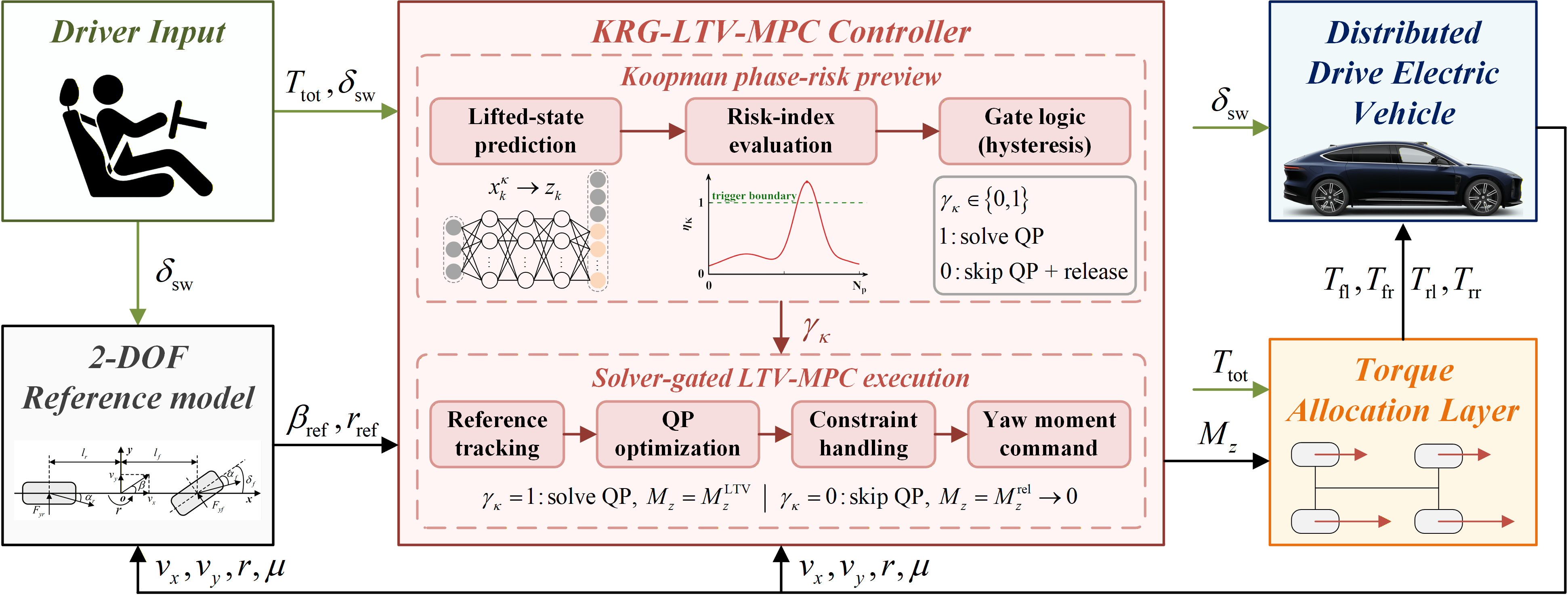}
	\caption{Overall architecture of the proposed KRG-LTV-MPC controller.}
	\label{fig:Overall_control_architecture}
\end{figure*}

The overall architecture of the proposed KRG-LTV-MPC controller is shown in Fig.~\ref{fig:Overall_control_architecture}. The Deep Koopman model provides a nominal phase-risk preview from the current vehicle state and driver inputs. The resulting risk signal determines the solver-gate state, while the LTV-MPC execution layer computes the physically constrained yaw moment only when the gate is active.
\subsection{Koopman phase-risk preview index}
\label{subsec:risk_preview}
A risk index based only on $|\beta|/\beta_{\lim}$ is insufficient for yaw stability preview. A large sideslip angle may still be acceptable if the state is moving back toward the stable origin, whereas a smaller sideslip angle may become risky if it is diverging rapidly. The proposed preview index therefore combines the predicted sideslip magnitude, the sideslip divergence tendency, and whether the yaw rate direction is favorable for reducing the sideslip angle.

The Koopman predictor provides a nominal preview of the sideslip angle and yaw rate over the prediction horizon $N_p^{\mathcal{K}}$. The predicted variables are normalized by the friction-adaptive stability envelope as
\begin{equation}
\left\{
\begin{aligned}
\hat{\beta}_{n,i}^{\mathcal{K}} &= \frac{\hat{\beta}_{k+i|k}^{\mathcal{K}}}{\beta_{\lim}},\\
\hat{r}_{n,i}^{\mathcal{K}} &= \frac{\hat{r}_{k+i|k}^{\mathcal{K}}}{r_{\lim}},
\end{aligned}
\right.
\qquad i=1,\ldots,N_p^{\mathcal{K}}.
\label{eq:koopman_norm_preview_final}
\end{equation}
The normalized sideslip rate term $\dot{\hat{\beta}}_{n,i}^{\mathcal{K}}$ is computed from the predicted sideslip sequence using a finite-difference operation. To distinguish divergence from convergence, two directional risk terms are defined as
\begin{equation}
\left\{
\begin{aligned}
d_{\beta,i}^{\mathcal{K}}
&=
\max\!\left(
0,\,
\operatorname{sgn}(\hat{\beta}_{n,i}^{\mathcal{K}})
\tau_{\beta}\dot{\hat{\beta}}_{n,i}^{\mathcal{K}}
\right),\\
d_{r,i}^{\mathcal{K}}
&=
\max\!\left(
0,\,
-\operatorname{sgn}(\hat{\beta}_{n,i}^{\mathcal{K}})\hat{r}_{n,i}^{\mathcal{K}}
\right),
\end{aligned}
\right.
\qquad i=1,\ldots,N_p^{\mathcal{K}}.
\label{eq:divergence_yaw_terms_final}
\end{equation}
Here, $d_{\beta,i}^{\mathcal{K}}$ becomes positive when the predicted sideslip tends to move away from zero, and $d_{r,i}^{\mathcal{K}}$ becomes positive when the predicted yaw rate direction is unfavorable for reducing the sideslip angle. The parameter $\tau_\beta$ is a time-scale factor that makes the normalized sideslip rate term dimensionless and controls the sensitivity of the risk index to sideslip divergence.

The Koopman phase-risk index is then constructed as
\begin{equation}
\left\{
\begin{aligned}
\chi_{\beta}^{\mathcal{K}}
&=
\max_{i=1,\ldots,N_p^{\mathcal{K}}}
\left|\hat{\beta}_{n,i}^{\mathcal{K}}\right|,\\
\chi_{\beta\dot{\beta}}^{\mathcal{K}}
&=
\max_{i=1,\ldots,N_p^{\mathcal{K}}}
\sqrt{
\left(\hat{\beta}_{n,i}^{\mathcal{K}}\right)^2
+\left(d_{\beta,i}^{\mathcal{K}}\right)^2
},\\
\chi_{\beta r}^{\mathcal{K}}
&=
\max_{i=1,\ldots,N_p^{\mathcal{K}}}
d_{r,i}^{\mathcal{K}},\\
\chi_{\mathcal{K}}
&=
\max\!\left(
 w_{\beta}\chi_{\beta}^{\mathcal{K}},\,
 w_{\beta\dot{\beta}}\chi_{\beta\dot{\beta}}^{\mathcal{K}},\,
 w_{\beta r}\chi_{\beta r}^{\mathcal{K}}
\right).
\end{aligned}
\right.
\label{eq:koopman_phase_risk_final}
\end{equation}
A weighted-maximum aggregation is adopted so that a single severe risk component can dominate the preview index instead of being diluted by the other components. The weights are set empirically according to a safety-oriented hierarchy. The largest weight is assigned to $\chi_{\beta\dot\beta}^{\mathcal K}$ because divergence away from the stable origin is more critical than a finite but bounded sideslip angle. The sideslip term $\chi_{\beta}^{\mathcal K}$ and the yaw-direction term $\chi_{\beta r}^{\mathcal K}$ are weighted more moderately to avoid excessive activation from bounded transient responses. The final values are given in Table~\ref{tab:controller_params}.

The bounded Koopman risk level used by the solver gate is
\begin{equation}
\ell_{\mathcal{K}}
=
\operatorname{sat}_{[0,1]}
\left(
\frac{\chi_{\mathcal{K}}-\zeta_0}{\zeta_1-\zeta_0}
\right),
\label{eq:koopman_risk_level_final}
\end{equation}
where $\zeta_0$ and $\zeta_1$ are the lower and upper breakpoints for risk-level normalization. The pair $(\chi_{\mathcal{K}},\ell_{\mathcal{K}})$ serves as the Koopman risk signal for the gate logic described in Section~\ref{sec:gated_control}.

\subsection{Solver-gated yaw moment execution}
\label{sec:gated_control}
The Koopman phase-risk preview is used as a solver-level supervisory gate for the LTV-MPC execution layer. At each sampling time, the Koopman risk index is evaluated before solving the LTV-MPC QP.

In addition to the Koopman preview risk, a current-state safeguard is defined as
\begin{equation}
	\chi_{\mathrm{cur}}
	=
	\max\!\left(
	|\beta_n|,\,
	w_{r,c}|r_n|
	\right),
	\label{eq:current_gate_risk_final}
\end{equation}
where $\beta_n=\beta/\beta_{\lim}$ and $r_n=r/r_{\lim}$. This term acts as a safeguard when the measured state is already close to the stability envelope.

The gate is switched on when at least one of the Koopman-preview or current-state risk branches reaches its activation threshold:
\begin{equation}
	\mathcal C_{\mathrm{on}}:
	\quad
	\chi_{\mathcal{K}}\ge \chi_{\mathrm{on}}^{\mathrm g}
	\ \lor\ 
	\ell_{\mathcal{K}}\ge \ell_{\mathrm{on}}^{\mathrm g}
	\ \lor\ 
	\chi_{\mathrm{cur}}\ge \chi_{\mathrm{cur,on}}^{\mathrm g}.
	\label{eq:gate_on_final}
\end{equation}
The on-condition must persist for $T_{\mathrm{on}}$ before the gate state is set to $\gamma_{\mathcal{K}}=1$. Conversely, the gate is switched off only when all risk branches fall below their off-thresholds:
\begin{equation}
	\mathcal C_{\mathrm{off}}:
	\quad
	\chi_{\mathcal{K}}\le \chi_{\mathrm{off}}^{\mathrm g}
	\ \land\ 
	\ell_{\mathcal{K}}\le \ell_{\mathrm{off}}^{\mathrm g}
	\ \land\ 
	\chi_{\mathrm{cur}}\le \chi_{\mathrm{cur,off}}^{\mathrm g}.
	\label{eq:gate_off_final}
\end{equation}
The off-condition must persist for $T_{\mathrm{off}}$ before the gate state is set to $\gamma_{\mathcal{K}}=0$. This hysteresis and persistence design avoids repeated activation and deactivation of the LTV-MPC QP during low-risk phases.

The final commanded yaw moment is determined by
\begin{equation}
	M_z(k)=
	\begin{cases}
		M_z^{\mathrm{LTV}}(k), & \gamma_{\mathcal{K}}(k)=1,\\
		M_z^{\mathrm{rel}}(k), & \gamma_{\mathcal{K}}(k)=0,
	\end{cases}
	\label{eq:gated_mz_final}
\end{equation}
where $\gamma_{\mathcal{K}}\in\{0,1\}$ is the Koopman risk-gate state. When
$\gamma_{\mathcal{K}}=1$, the LTV-MPC QP is solved and the
resulting additional yaw moment $M_z^{\mathrm{LTV}}$ is applied. When $\gamma_{\mathcal{K}}=0$, the QP is skipped and the previously commanded additional yaw moment is tapered to zero according to
\begin{equation}
	M_z^{\mathrm{rel}}(k)
	=
	\operatorname{sgn}\!\left(M_z(k-1)\right)
	\max\!\left(
	|M_z(k-1)|-\Delta M_{z,\mathrm{rel}},
	0
	\right).
	\label{eq:mz_release_final}
\end{equation}
The release rate is chosen to be no larger than the increment bound of the yaw moment used in the LTV-MPC execution layer, i.e. $\Delta M_{z,\mathrm{rel}}\leq\Delta M_{z,\max}$, so that the off-mode command does not introduce a discontinuous step of the yaw moment.

Note that the Koopman predictor generates nominal risk preview without explicitly accounting for the additional yaw moment. After DYC has suppressed the sideslip response, the predictor is initialized from the corrected state and may underestimate the risk that would reappear if the yaw moment were removed abruptly. This effect is limited by the off-persistence time $T_{\rm off}$ and the bounded-rate yaw moment release in Eq.~\eqref{eq:mz_release_final}.
\section{Validation results and discussion}
\label{sec:experiments}
\subsection{Simulation setup and evaluation metrics}
\label{subsec:simulation_platform}

\begin{table}[!t] 
	\centering 
    \footnotesize
	\caption{Parameters of vehicle and in-wheel motor.}
	\label{tab:vehicle_motor_params}
	\begin{tabular*}{\columnwidth}{@{\extracolsep{\fill}} llll @{}}
		\toprule
		Symbol & Value & Unit & Description \\
		\midrule
		\multicolumn{4}{@{}l}{\emph{Vehicle}} \\
		$m$    & 1525.1 & kg & Vehicle mass \\
		$I_z$  & 2315.3 & $\text{kg}\cdot\text{m}^2$ & Yaw moment of inertia \\
		$l_f$  & 1.110  & m  & Distance from CG to front axle \\
		$l_r$  & 1.756  & m  & Distance from CG to rear axle \\
		$C_{\alpha f}$ & 162615 & N/rad & Cornering stiffness of front axle \\
		$C_{\alpha r}$ & 112440 & N/rad & Cornering stiffness of rear axle \\
		$B_f$  & 1.550  & m  & Front track width \\
		$B_r$  & 1.550  & m  & Rear track width \\
		$R_w$  & 0.325  & m  & Effective wheel radius \\
		\midrule
		\multicolumn{4}{@{}l}{\emph{In-wheel motor}} \\
		$T_{\max}$      & 1250 & Nm  & Single-motor peak torque \\
		$P_{\max}$      & 80   & kW  & Single-motor peak power \\
		$\omega_{\max}$ & 1600 & rpm & Maximum motor speed \\
		\bottomrule
	\end{tabular*}
\end{table}

\begin{table}[!t]
	\centering
    \footnotesize
	\setlength{\tabcolsep}{3pt} 
	\caption{Parameters of the LTV-MPC baseline and Koopman risk gate.}
	\label{tab:controller_params}
	\begin{tabular*}{\columnwidth}{@{\extracolsep{\fill}} lll @{}}
		\toprule
		Parameter & Value & Description \\
		\midrule
		\multicolumn{3}{@{}l}{\emph{LTV-MPC}} \\
		$N_p$ & 20 & Prediction horizon \\
		$N_c$ & 20 & Control horizon \\
		$Q_\beta$ & $1.0\times10^5$ & Sideslip weight \\
		$Q_r$ & $1.0\times10^4$ & Yaw rate weight \\
		$R_M$ & $5.0\times10^{-4}$ & Moment penalty \\
		$R_{\Delta M}$ & $1.0\times10^{-1}$ & Moment rate penalty \\
		$\rho_{\epsilon}$ & $3.0\times10^{6}$ & Slack penalty \\
		$M_{z,\max}$ & 3000 Nm & Moment limit \\
		$\Delta M_{z,\max}$ & 500 Nm/sample & Rate limit \\
		$\epsilon_{\max}$ & 10 & Upper bound of slack \\
		\midrule
		\multicolumn{3}{@{}l}{\emph{Koopman risk preview}} \\
		$N_p^{\mathcal{K}}$ & 50 & Preview horizon \\
		$\tau_\beta$ & 0.20 s & Sideslip-divergence scaling time \\
		$w_\beta,w_{\beta\dot\beta},w_{\beta r}$ & 0.20, 1.00, 0.25 & Preview-risk weights \\
		$w_{r,c}$ & 0.20 & Current-state yaw rate weight \\
		$\zeta_0,\zeta_1$ & 0.75, 1.10 & Risk-level normalization bounds \\
		\midrule
		\multicolumn{3}{@{}l}{\emph{Koopman risk gate: high-$\mu$}} \\
		$\chi_{\rm on}^{\rm g},\chi_{\rm off}^{\rm g}$ & 0.36, 0.20 & Preview-risk thresholds \\
		$\ell_{\rm on}^{\rm g},\ell_{\rm off}^{\rm g}$ & 0.20, 0.05 & Risk-level thresholds \\
		$\chi_{\rm cur,on}^{\rm g},\chi_{\rm cur,off}^{\rm g}$ & 0.35, 0.18 & Current-risk thresholds \\
		$T_{\rm on},T_{\rm off}$ & 0.02 s, 0.45 s & Gate persistence \\
		\midrule
		\multicolumn{3}{@{}l}{\emph{Koopman risk gate: low-$\mu$}} \\
		$\chi_{\rm on}^{\rm g},\chi_{\rm off}^{\rm g}$ & 1.00, 0.75 & Preview-risk thresholds \\
		$\ell_{\rm on}^{\rm g},\ell_{\rm off}^{\rm g}$ & 0.60, 0.20 & Risk-level thresholds \\
		$\chi_{\rm cur,on}^{\rm g},\chi_{\rm cur,off}^{\rm g}$ & 0.55, 0.30 & Current-risk thresholds \\
		$T_{\rm on},T_{\rm off}$ & 0.03 s, 0.50 s & Gate persistence \\
		\bottomrule
	\end{tabular*}
\end{table}

The proposed method is validated through co-simulation between CarSim 2024 and MATLAB R2025b, executed on a laptop equipped with an AMD Ryzen 9 9955HX CPU. The controller layer runs at a sampling time of $T_s=0.01$~s. All double lane change (DLC) trajectories are generated using the CarSim Double Lane Change (Quick Start) maneuver (ISO 3888-1), with the built-in preview driver model providing steering inputs. The QP is solved by the active-set algorithm of the MATLAB quadprog function. The vehicle and motor parameters are listed in Table~\ref{tab:vehicle_motor_params}, and the controller parameters are summarized in Table~\ref{tab:controller_params}. The high- and low-friction gate thresholds in Table~\ref{tab:controller_params} were selected by offline calibration on representative DLC maneuvers to balance early activation in critical conditions and reduced QP solving during low-risk intervals. The parameter sets are switched online according to the road-friction coefficient provided by CarSim. Three controllers are compared:
\begin{enumerate}
    \item \textbf{No-DYC}: no additional yaw moment is applied.
    \item \textbf{LTV-MPC}: baseline; QP is solved at every sample.
    \item \textbf{KRG-LTV-MPC}: proposed; QP solved only when the
        Koopman risk gate is active, with bounded-rate moment release otherwise.
\end{enumerate}

Performance is assessed by the lateral tracking RMS error $e_{y,\rm rms}$, the peak sideslip angle $|\beta|_{\max}$, the RMS and peak steering wheel angles $\delta_{\mathrm{sw,rms}}$ and $|\delta_{\mathrm{sw}}|_{\max}$, the peak additional yaw moment $|M_z|_{\max}$, and the cumulative additional yaw moment intervention index
\begin{equation}
    J_{M_z}=\int_0^{T_{\mathrm{sim}}}|M_z (t)|\,dt,
    \label{eq:JMZ_final}
\end{equation}
where $T_{\mathrm{sim}}$ is the full duration of the simulation. The relative reduction $\Delta J_{M_z}$ is computed with respect to the LTV-MPC baseline, and $D_{\rm QP}$ denotes the percentage of samples over the full simulation horizon for which the QP is solved.

\subsection{Shadow-mode phase-risk preview}
\label{subsec:risk_results}

\begin{figure}
	\centering
	\includegraphics[width=\columnwidth]{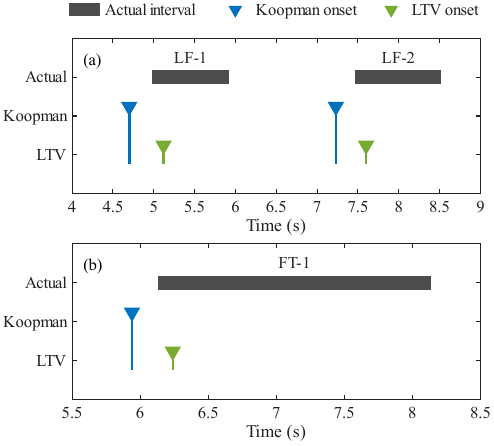}
    \caption{Shadow-mode phase-risk preview timing. (a) Low-friction case. (b) Friction-transition case. Gray bars denote labeled actual phase-risk intervals, and colored markers denote event-associated warning onsets.}
	\label{fig:shadowmode_event_preview}
\end{figure}

\begin{table}[!t]
	\centering
	\footnotesize
	\setlength{\tabcolsep}{3pt}
	\caption{Event-level shadow-mode lead-time results. LF denotes DLC 72~km/h with $\mu=0.40$, and FT denotes DLC 90~km/h with $\mu=0.85\rightarrow0.40$.}
	\label{tab:risk_preview_results}
	\begin{tabular*}{\columnwidth}{@{\extracolsep{\fill}} lclcc @{}}
		\toprule
		Event ID & Actual interval & Predictor & Warning onset & $\Delta t_{\rm lead}$ \\
		\midrule
		\multirow{2}{*}{LF-1} 
		& \multirow{2}{*}{4.98--5.92 s} & Koopman & 4.70 s & $+$0.28 s \\
		&    & LTV            & 5.12 s & $-$0.14 s \\
		\addlinespace
		\multirow{2}{*}{LF-2}
		& \multirow{2}{*}{7.47--8.52 s} & Koopman  & 7.23 s & $+$0.24 s \\
		&    & LTV            & 7.60 s & $-$0.13 s \\
		\addlinespace
		\multirow{2}{*}{FT-1}
		& \multirow{2}{*}{6.13--8.14 s} & Koopman  & 5.94 s & $+$0.19 s \\
		&    & LTV            & 6.24 s & $-$0.11 s \\
		\bottomrule
	\end{tabular*}
\end{table}

The first experiment isolates the risk-preview capability from the closed-loop control action. The Koopman and LTV predictors are evaluated in shadow mode along measured No-DYC trajectories under the low-friction and friction-transition DLC conditions, using the current measured inputs held constant over the prediction horizon. For event-level timing evaluation, a unit phase-risk boundary is used to identify actual intervals and warning onsets, separately from the friction-adaptive thresholds used for closed-loop solver activation. The labeled actual intervals are obtained offline from the measured No-DYC phase-risk response and used only as timing references. Since the high-friction No-DYC response remains within the stability envelope, the lead-time evaluation is restricted to the low-friction and friction-transition cases.

The key metric is the timing difference between the event-associated warning onset and the actual phase-risk onset. As shown in Fig.~\ref{fig:shadowmode_event_preview} and Table~\ref{tab:risk_preview_results}, the Koopman predictor reaches the warning boundary before the measured phase-risk onset in all three evaluated events, whereas the LTV predictor reaches the warning boundary after the event has already started. The Koopman predictor achieves positive lead times across all three evaluated events: 0.28~s and 0.24~s under low friction, and 0.19~s under friction transition. By contrast, the LTV predictor lags behind the actual onset by 0.14~s, 0.13~s, and 0.11~s, respectively. These event-level results support the use of the Koopman predictor as a preview-oriented phase-risk detector under critical low-friction and friction-transition conditions.

\subsection{Closed-loop stability and intervention analysis}
Three DLC scenarios are evaluated: high-friction ($\mu=0.85$, 90~km/h), low-friction ($\mu=0.40$, 72~km/h), and a friction-transition ($\mu=0.85\rightarrow0.40$, 90~km/h). The quantitative results are summarized in Table~\ref{tab:closed_loop_summary}.

\begin{table*}[!t]
	\centering
	\caption{Closed-loop performance and DYC intervention comparison under three DLC conditions.}
	\label{tab:closed_loop_summary}
	\begin{tabular*}{\textwidth}{@{\extracolsep{\fill}} llllllllll @{}}
		\toprule
		\multirow{3}{*}{Condition}
		& \multirow{3}{*}{Controller}
		& \multicolumn{2}{c}{Closed-loop response}
		& \multicolumn{2}{c}{Steering demand}
		& \multicolumn{4}{c}{DYC intervention} \\
		\cmidrule(lr){3-4} \cmidrule(lr){5-6} \cmidrule(lr){7-10} 
		& & $e_{y,\mathrm{rms}}$ & $|\beta|_{\max}$ & $\delta_{\mathrm{sw,rms}}$ & $|\delta_{\mathrm{sw}}|_{\max}$ & $|M_z|_{\max}$ & $J_{M_z}$ & $\Delta J_{M_z}$ & $D_{\rm QP}$ \\
		& & (m) & (deg) & (deg) & (deg) & (kNm) & (kNm$\cdot$s) & (\%) & (\%) \\
		\midrule
		\multirow{3}{*}{\makecell[l]{DLC 90 km/h \\ $\mu=0.85$}}
		& No-DYC      & 0.175 & 3.095 & 27.243 & 77.729 & 0     & 0     & --    & --  \\
		& LTV-MPC     & 0.177 & 2.366 & 27.679 & 77.218 & 1.265 & 1.453 & 0     & 100.0  \\
		& KRG-LTV-MPC & 0.175 & 2.601 & 27.258 & 74.754 & 1.095 & 1.093 & 24.8  & 32.8  \\
		\midrule
		\multirow{3}{*}{\makecell[l]{DLC 72 km/h \\ $\mu=0.40$}}
		& No-DYC      & 0.219 & 10.799 & 66.761 & 175.503 & 0     & 0     & --    & --   \\
		& LTV-MPC     & 0.279 & 1.074  & 41.001 & 112.657 & 1.923 & 3.020 & 0     & 100.0 \\
		& KRG-LTV-MPC & 0.242 & 2.105  & 37.698 & 98.706  & 1.364 & 1.695 & 43.9  & 41.0 \\
		\midrule
		\multirow{3}{*}{\makecell[l]{DLC 90 km/h \\ $\mu=0.85\rightarrow0.40$}}
		& No-DYC      & 0.285 & 12.456 & 63.861 & 200.494 & 0     & 0     & --    & --   \\
		& LTV-MPC     & 0.324 & 2.680  & 35.061 & 116.772 & 1.687 & 1.664 & 0     & 100.0  \\
		& KRG-LTV-MPC & 0.306 & 2.801  & 32.851 & 105.447 & 1.486 & 1.241 & 25.4  & 42.0 \\
		\bottomrule
	\end{tabular*}
\end{table*}

\subsubsection{High-friction DLC}
\label{subsec:closed_loop_high_mu}
\begin{figure}
	\centering
	\includegraphics[width=\columnwidth]{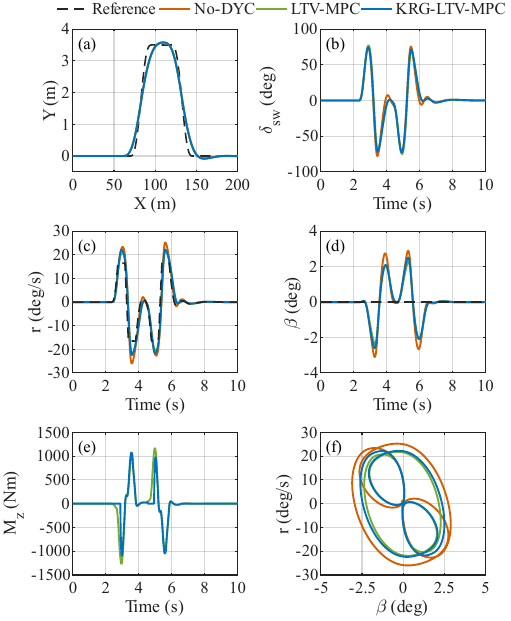}
	\caption{Closed-loop responses under the DLC maneuver at 90 km/h and $\mu=0.85$.
		(a) Vehicle trajectory. (b) Steering wheel angle. (c) Yaw rate. (d) Sideslip angle. (e) Additional yaw moment. (f) Sideslip--yaw rate phase portrait.}
	\label{fig:dlc90_mu085_closedloop}
\end{figure}

The high-friction DLC case evaluates the capability of the proposed supervisor to avoid continuous stabilizing interventions when the vehicle operates within the stable envelope. As shown in Fig.~\ref{fig:dlc90_mu085_closedloop}, all three controllers maintain comparable path-following behavior, and the No-DYC vehicle does not develop a severe phase plane excursion. In this regime, DYC refines the transient sideslip and yaw rate response during high-speed steering rather than recovering the vehicle from instability. The advantage of KRG-LTV-MPC lies not in tracking improvement, but in preserving a response close to the LTV-MPC while avoiding continuous QP solving. Quantitatively, KRG-LTV-MPC solves the QP for only $32.8\%$ of the samples, indicating that the Koopman gate avoids continuous yaw moment regulation when the predicted state risk is low. In terms of control performance, although the reduction in $|M_z|_{\max}$ (from $1.265$ to $1.095$~kNm) is modest, the $24.8\%$ reduction in $J_{M_z}$ reflects a faster return of the commanded moment to zero after each steering-reversal event.

\subsubsection{Low-friction DLC}
\label{subsec:closed_loop_low_mu}
\begin{figure}
	\centering
    \includegraphics[width=\columnwidth]{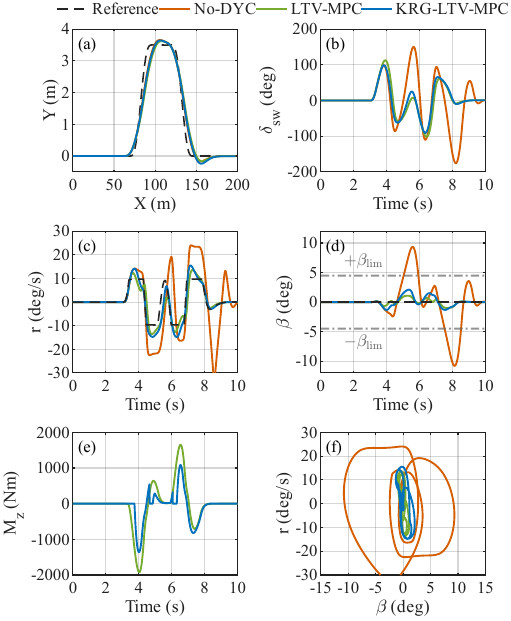}
	\caption{Closed-loop responses under the DLC maneuver at 72 km/h and $\mu=0.4$.
		(a) Vehicle trajectory. (b) Steering wheel angle. (c) Yaw rate. (d) Sideslip angle. (e) Additional yaw moment. (f) Sideslip--yaw rate phase portrait.}
	\label{fig:dlc72_mu04_closedloop}
\end{figure}

The low-friction DLC case is the most critical test of the proposed low-intervention strategy. Fig.~\ref{fig:dlc72_mu04_closedloop} shows that the vehicle without DYC suffers from a large sideslip excursion and a diverging phase trajectory, indicating that steering correction alone is insufficient to maintain yaw stability. The LTV-MPC gives the tightest sideslip regulation, but it does so through continuous and strong yaw moment intervention. KRG-LTV-MPC accepts a slightly larger peak sideslip angle than LTV-MPC, but maintains both $\beta$ and $r$ within the stability boundary defined in Eq.~\eqref{eq:basic_envelope_final} while substantially reducing intervention demand.

Quantitatively, KRG-LTV-MPC reduces $J_{M_z}$ from $3.020$ to $1.695~{\rm kNm\,s}$ and $|M_z|_{\max}$ from $1.923$ to $1.364~{\rm kNm}$ relative to LTV-MPC. The steering demand is also reduced, with both $\delta_{\rm sw,rms}$ and $|\delta_{\rm sw}|_{\max}$ lower than those of LTV-MPC. Although LTV-MPC gives the smallest $|\beta|_{\max}$, it produces a larger $e_{y,\rm rms}$ because the execution layer prioritizes $\beta$ and $r$ regulation rather than direct path-tracking error minimization. By tapering the yaw moment during low-risk intervals, KRG-LTV-MPC reduces persistent interaction with the driver model and yields a lower $e_{y,\rm rms}$ while keeping both $\beta$ and $r$ within the stability boundary. Thus, the main finding in the low-friction case is not that KRG-LTV-MPC minimizes sideslip more aggressively than LTV-MPC, but that it maintains an acceptable stability margin with a 43.9\% reduction in cumulative yaw moment intervention and a QP solve ratio of 41.0\%.

\subsubsection{Friction-transition DLC}
\label{subsec:closed_loop_mu_jump}
\begin{figure}
	\centering
	\includegraphics[width=\columnwidth]{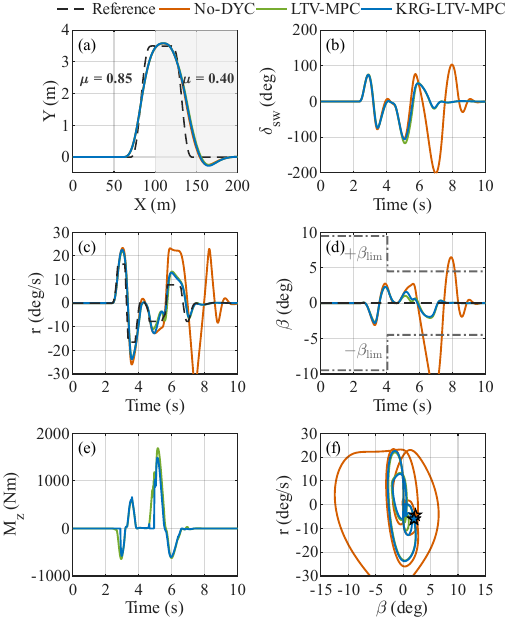}
	\caption{Closed-loop responses under the DLC maneuver at 90 km/h with a road-friction transition from $\mu=0.85$ to $\mu=0.4$.
		(a) Vehicle trajectory. (b) Steering wheel angle. (c) Yaw rate. (d) Sideslip angle. (e) Additional yaw moment. (f) Sideslip--yaw rate phase portrait. The star markers in (f) indicate the states at the friction-transition instant.}
	\label{fig:dlc90_mu085to04_closedloop}
\end{figure}
The friction-transition DLC evaluates whether the gate can respond to a sudden increase in instability risk rather than relying on a fixed always-on intervention policy. As shown in Fig.~\ref{fig:dlc90_mu085to04_closedloop}, after the road-friction transition, the No-DYC vehicle exhibits a rapid growth of sideslip angle and a large phase plane excursion. Both DYC controllers suppress this growth, but KRG-LTV-MPC achieves a response close to LTV-MPC with lower intervention intensity. Specifically, $|\beta|_{\max}$ is 2.801$^{\circ}$ for KRG-LTV-MPC and 2.680$^{\circ}$ for LTV-MPC, while $J_{M_z}$ is reduced by 25.4\%. The result indicates that the Koopman gate preserves the stabilizing effect of the LTV-MPC execution layer during the friction drop, but avoids the sustained yaw-moment level that LTV-MPC maintains throughout the entire maneuver, concentrating intervention in the risk-relevant phase following the friction drop.

\begin{figure}
	\centering
	\includegraphics[width=\columnwidth]{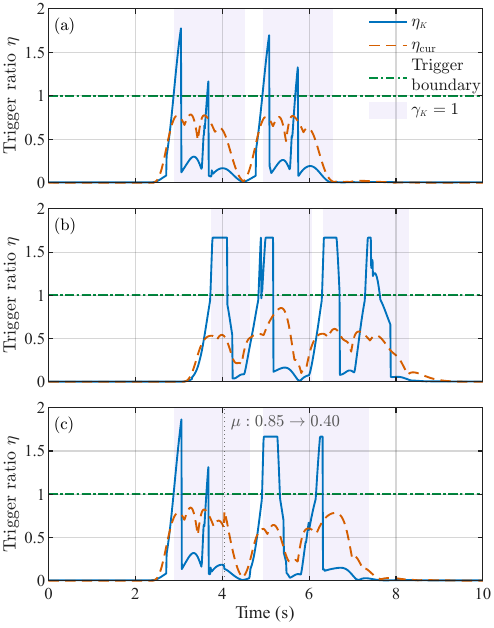}
	\caption{Koopman risk-gated activation behavior under three DLC conditions. (a) 90 km/h, $\mu=0.85$. (b) 72 km/h, $\mu=0.40$. (c) 90 km/h, $\mu=0.85\rightarrow0.40$. The gray vertical dotted line in (c) indicates the prescribed road-friction drop.}
	\label{fig:krg_gate_activation}
\end{figure}
To explain the intervention reduction observed in the closed-loop cases, Fig.~\ref{fig:krg_gate_activation} reports the normalized trigger ratios. The Koopman trigger ratio is defined as $\eta_{\mathcal{K}}=\max(\chi_{\mathcal{K}}/\chi_{\rm on}^{\rm g},\ell_{\mathcal{K}}/\ell_{\rm on}^{\rm g})$, and the current-state trigger ratio is $\eta_{\rm cur}=\chi_{\rm cur}/\chi_{\rm cur,on}^{\rm g}$. The activation boundary corresponds to $\eta=1$. In all three DLC conditions, $\eta_{\mathcal{K}}$ dominates QP activation, while $\eta_{\rm cur}$ remains below unity. This indicates that the Koopman preview captures the growth of predicted phase risk before the measured current-state risk reaches the activation threshold, consistent with the positive warning lead times observed in the shadow-mode evaluation. The resulting gate activation is therefore concentrated in the lane-change portions of the DLC maneuver rather than over the full simulation horizon. In the friction-transition case, the gate parameter set switches from high-$\mu$ to low-$\mu$ values at the marked friction-transition instant in Fig.~\ref{fig:krg_gate_activation}(c), consistent with the friction-adaptive stability envelope in Eq.~\eqref{eq:lim_final}.

The shaded regions are determined by the hysteretic gate state $\gamma_{\mathcal{K}}$ rather than by instantaneous threshold crossings of $\eta_{\mathcal{K}}$ alone. Hence, the gate can remain active after $\eta_{\mathcal K}$ falls below unity until the off-condition is satisfied. The full-horizon QP solve ratios remain substantially below the always-on LTV-MPC baseline, with $D_{\rm QP}=32.8\%$, $41.0\%$, and $42.0\%$ for the three DLC conditions.

Overall, the reductions in $J_{M_z}$, $|M_z|_{\max}$, steering demand, and $D_{\rm QP}$ indicate that the proposed method acts as a low-intervention yaw stability supervisor rather than as a controller designed to minimize sideslip at all times.
\subsection{Gate-source ablation}
\label{subsec:gate_ablation}
To isolate the value of Koopman-lifted phase-risk preview from that of the DYC execution law, two gated baselines are introduced that preserve the solver-gated mechanism while modifying only the risk-source input. The first baseline, denoted as Cur-gated, removes the preview branch and activates the LTV-MPC execution layer only according to the measured current-state risk. The second baseline, denoted as LTV-gated, replaces the Koopman preview risk with the phase-risk index computed from the local LTV prediction. All gated controllers retain the same LTV-MPC execution layer, yaw moment constraints, hysteresis logic, and release mechanism.

\begin{table*}[!t] 
	\centering
	\caption{Ablation results of different gate sources.}
	\label{tab:gate_source_ablation}
	\begin{tabular*}{\textwidth}{@{\extracolsep{\fill}} lllllll @{}}
		\toprule
		Condition & Controller &
		\makecell[c]{$|\beta|_{\max}$ \\ (deg)} &
		\makecell[c]{$\delta_{\rm sw,rms}$ \\ (deg)} &
		\makecell[c]{$|M_z|_{\max}$ \\ (kNm)} &
		\makecell[c]{$J_{M_z}$ \\ (kNm$\cdot$s)} &
		\makecell[c]{$D_{\rm QP}$ \\ (\%)} \\
		\midrule
		\multirow{4}{*}{\makecell[l]{DLC 72 km/h \\ $\mu=0.40$}}
		& LTV-MPC & 1.074 & 41.001 & 1.923 & 3.020 & 100.0 \\
		& Cur-gated & 24.168 & 172.527 & 2.999 & 8.538 & 51.4 \\
		& LTV-gated & 2.905 & 39.945 & 1.317 & 1.504 & 35.7 \\
		& KRG-LTV-MPC & 2.105 & 37.698 & 1.364 & 1.695 & 41.0 \\
		\midrule
		\multirow{4}{*}{\makecell[l]{DLC 90 km/h \\ $\mu=0.85\rightarrow0.40$}}
		& LTV-MPC & 2.680 & 35.061 & 1.687 & 1.664 & 100.0 \\
		& Cur-gated & 6.322 & 41.543 & 1.536 & 1.672 & 29.2 \\
		& LTV-gated & 2.681 & 33.860 & 1.555 & 1.460 & 44.8 \\
		& KRG-LTV-MPC & 2.801 & 32.851 & 1.486 & 1.241 & 42.0 \\
		\bottomrule
	\end{tabular*}
\end{table*}

\begin{figure}
	\centering
	\includegraphics[width=\columnwidth]{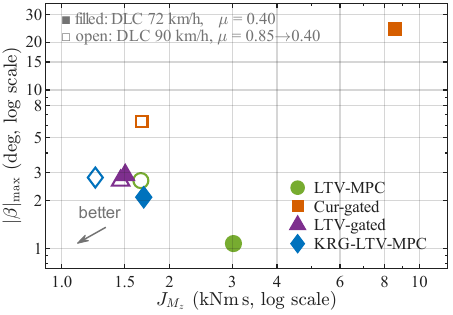}
	\caption{Stability--intervention trade-off of different gate sources.}
	\label{fig:trade-off}
\end{figure}
Fig.~\ref{fig:trade-off} visualizes the stability--intervention trade-off using $J_{M_z}$ and $|\beta|_{\max}$, where points closer to the lower-left region indicate lower yaw moment effort and smaller sideslip excursions. The Cur-gated points are displaced away from this favorable region in both critical conditions, indicating that current-state risk alone is not a reliable gate source. Table~\ref{tab:gate_source_ablation} confirms this trend: in the low-friction case, Cur-gated reacts only after a large phase plane excursion has developed, resulting in the largest $|\beta|_{\max}$ ($24.168^\circ$) and $J_{M_z}$ ($8.538~{\rm kNm\,s}$) among all controllers. In the friction-transition case, it achieves the lowest $D_{\rm QP}$ ($29.2\%$), but at the cost of a much larger $|\beta|_{\max}$ ($6.322^\circ$). Thus, Cur-gated gives an unreliable stability--intervention trade-off under critical friction conditions.

The comparison between LTV-gated and KRG-LTV-MPC further shows the role of Koopman-lifted phase-risk preview. In Fig.~\ref{fig:trade-off}, LTV-gated moves leftward relative to KRG-LTV-MPC in the low-friction case, but also upward, indicating lower yaw moment effort at the cost of looser sideslip containment. In the friction-transition case, KRG-LTV-MPC remains at a similar sideslip level while shifting leftward relative to LTV-gated. The numerical values in Table~\ref{tab:gate_source_ablation} support this interpretation: LTV-gated reduces $J_{M_z}$ to $1.504~{\rm kNm\,s}$ under low friction but increases $|\beta|_{\max}$ to $2.905^\circ$, while KRG-LTV-MPC limits $|\beta|_{\max}$ to $2.105^\circ$. Under friction transition, KRG-LTV-MPC reduces $J_{M_z}$ from $1.460$ to $1.241~{\rm kNm\,s}$ and decreases $D_{\rm QP}$ from $44.8\%$ to $42.0\%$ while maintaining comparable sideslip containment. These results indicate that the Koopman gate provides a more consistent stability--intervention trade-off across the two critical friction conditions.

Overall, the ablation results show that reducing the QP solve ratio alone is insufficient. Cur-gated activates too late, whereas LTV-gated gives a less consistent stability--intervention trade-off across friction regimes. These results support the use of Koopman-lifted phase-risk preview for low-intervention DYC.
\subsection{Computational efficiency}
\label{subsec:computational_efficiency}
\begin{figure}
	\centering
	\includegraphics[width=\columnwidth]{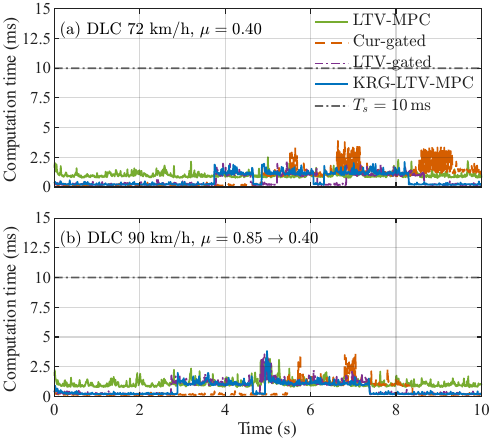}
	\caption{Single-step computation time of the compared controllers.}
	\label{fig:computation_time}
\end{figure}

\begin{table}[!t]
	\centering
	\setlength{\tabcolsep}{3pt}
	\caption{Comparison of computational performance.}
	\label{tab:computational_time}
	\begin{tabular*}{\columnwidth}{@{\extracolsep{\fill}} lllll @{}}
		\toprule
		Condition & Controller & \makecell[c]{Mean \\ (ms)} & \makecell[c]{Max \\ (ms)} & \makecell[c]{$D_{\rm QP}$ \\ (\%)} \\
		\midrule
		\multirow{4}{*}{\makecell[l]{DLC 72 km/h \\ $\mu=0.40$}}
		& LTV-MPC & 1.010 & 2.510 & 100.0 \\
		& Cur-gated & 0.826 & 3.883 & 51.4\\
		& LTV-gated & 0.542 & 2.218 & 35.7\\
		& KRG-LTV-MPC & 0.641 & 2.108 & 41.0\\
		\midrule
		\multirow{4}{*}{\makecell[l]{DLC 90 km/h \\ $\mu=0.85\rightarrow0.40$}}
		& LTV-MPC & 1.087 & 3.234 & 100.0\\
		& Cur-gated & 0.485 & 3.628 & 29.2\\
		& LTV-gated & 0.668 & 3.547 & 44.8\\
		& KRG-LTV-MPC & 0.623 & 3.816 & 42.0\\
		\bottomrule
	\end{tabular*}
\end{table}

Fig.~\ref{fig:computation_time} and Table~\ref{tab:computational_time} summarize the single-step computation time under the two critical DLC conditions. The reported time covers the entire controller block, including risk-preview calculation, gate update, and the LTV-MPC QP when applicable. The sampling period is $T_s=10~{\rm ms}$.

As shown in Fig.~\ref{fig:computation_time}, all controllers remain well below the sampling period throughout both maneuvers. The LTV-MPC maintains a continuous computational load because the QP is solved at every step, whereas the gated controllers exhibit a low/high computation-time pattern: lower values correspond to gate-off intervals where the QP is skipped, and higher values correspond to gate-on intervals where the QP is solved. For KRG-LTV-MPC, the mean computation time decreases from $1.010$ to $0.641$~ms in the low-friction case and from $1.087$ to $0.623$~ms in the friction-transition case, with corresponding QP solve ratios of $41.0\%$ and $42.0\%$. Its maximum computation times, $2.108$ and $3.816$~ms, also remain below the sampling period.

It is worth noting that, in the low-friction case, KRG-LTV-MPC achieves both a lower mean time and a lower maximum time than Cur-gated, while also avoiding the large sideslip excursion reported in Table~\ref{tab:gate_source_ablation}. Although LTV-gated gives the lowest mean computation time in the low-friction case ($0.542$~ms) and Cur-gated gives the lowest mean computation time in the friction-transition case ($0.485$~ms), the ablation results in Section~\ref{subsec:gate_ablation} show that these lower timings are associated with looser sideslip containment or delayed intervention. Therefore, the computational results of KRG-LTV-MPC should be interpreted together with its closed-loop stability and intervention performance. These timings were measured on a laptop, and embedded real-time feasibility still requires further verification.
\section{Conclusion}
\label{sec:conclusion}
This paper proposed a KRG-LTV-MPC framework for low-intervention DYC of DDEVs. The central idea is to separate Koopman-based phase-risk preview from safety-critical yaw moment execution: the Deep Koopman model anticipates the future sideslip--yaw rate risk evolution, while the LTV-MPC layer retains responsibility for physically constrained yaw moment generation. The results show that this role-separated architecture enables earlier phase-risk detection than the local LTV predictor and reduces unnecessary QP solving and yaw moment intervention without sacrificing yaw stability. The reduced steering demand further indicates that the intervention saving is not simply transferred to the driver model. Gate-source ablation confirms that the benefit does not come from solver gating alone. Cur-gated activates too late under critical conditions, while LTV-gated gives a less consistent stability--intervention trade-off across friction regimes. The proposed KRG-LTV-MPC provides a more consistent stability--intervention trade-off and reduces the average computation time while keeping all maximum computation times within the 10~ms sampling period. Overall, these results support using Koopman-based phase-risk preview as a supervisory layer for low-intervention DYC, not as an optimizer-internal execution model.

Future work will focus on experimental validation using a scaled vehicle platform and on controlled Koopman formulations.

\section*{DECLARATIONS}

\subsection*{Conflict of Interest}
The authors declare that they have no known competing financial interests or personal relationships that could have appeared to influence the work reported in this paper.

\subsection*{Authors' Contributions}
Wenjie Wang: Conceptualization, Methodology, Validation, Formal analysis, Writing-original draft, Writing-review \& editing. Hao Chen: Conceptualization, Validation, Writing-review \& editing. Ran Shu: Investigation, Writing-review \& editing. Kyoungseok Han: Resources, Writing-review \& editing. Hongyu Shu: Funding acquisition, Supervision.

\subsection*{Funding}
This work was supported by the National Natural Science Foundation of China (No. 52372376), the Graduate Research and Innovation Foundation of Chongqing (No. CYB240012), and the China Scholarship Council (No. 202506050050).

\subsection*{Data Availability}
Data will be made available on request.

\bibliographystyle{unsrt}
\bibliography{refs}

\end{document}